\documentclass[conference]{IEEEtran}
\IEEEoverridecommandlockouts
\usepackage{algorithm} 
\usepackage{algorithm}
\usepackage{algorithmic}
\usepackage{cite}
\usepackage{array}
\usepackage{amsmath,amssymb,amsfonts}
\usepackage{algorithmic}
\usepackage{booktabs} 
\usepackage{makecell} 
\usepackage{graphicx}  
\usepackage{subcaption} 
\usepackage{caption}
\usepackage{textcomp}
\usepackage{xcolor}
\usepackage{colortbl} 
\usepackage{wrapfig}
\def\BibTeX{{\rm B\kern-.05em{\sc i\kern-.025em b}\kern-.08em
    T\kern-.1667em\lower.7ex\hbox{E}\kern-.125emX}}
\begin{document}

\title{Maximizing User Connectivity in AI-Enabled Multi-UAV Networks: A Distributed Strategy Generalized to Arbitrary User Distributions
}

\author{\IEEEauthorblockN{
Bowei Li\IEEEauthorrefmark{1},
Yang Xu\IEEEauthorrefmark{2},
Ran Zhang\IEEEauthorrefmark{2},
Jiang (Linda) Xie\IEEEauthorrefmark{2},
and Miao Wang\IEEEauthorrefmark{2}} \\
\IEEEauthorblockA{\IEEEauthorrefmark{1}
Department of Electrical and Computer Engineering, Carnegie Mellon University, Pittsburgh, PA, USA}
\IEEEauthorblockA{\IEEEauthorrefmark{2}
Department of Electrical and Computer Engineering, 
University of North Carolina at Charlotte, Charlotte, NC, USA}
\IEEEauthorblockA{Email: \IEEEauthorrefmark{1}$boweili$@andrew.cmu.edu,\IEEEauthorrefmark{2}\{$yxu33, rzhang8, jxie1, mwang25$\}@charlotte.edu}
}

\maketitle

\begin{abstract}
Deep reinforcement learning (DRL) has been extensively applied to Multi-Unmanned Aerial Vehicle (UAV) network (MUN) to effectively enable real-time adaptation to complex, time-varying environments. Nevertheless, most of the existing works assume a stationary user distribution (UD) or a dynamic one with predicted patterns. Such considerations may make the UD-specific strategies insufficient when a MUN is deployed in unknown environments. To this end,
this paper investigates distributed user connectivity maximization problem in a MUN with generalization to arbitrary UDs. Specifically, the problem is first formulated into a time-coupled combinatorial nonlinear non-convex optimization with arbitrary underlying UDs. To make the optimization tractable, a multi-agent CNN-enhanced deep Q learning (MA-CDQL) algorithm is proposed. The algorithm integrates a ResNet-based CNN to the policy network to analyze the input UD in real time and obtain optimal decisions based on the extracted high-level UD features. To improve the learning efficiency and avoid local optimums, a heatmap algorithm is developed to transform the raw UD to a continuous density map. The map will be part of the true input to the policy network. Simulations are conducted to demonstrate the efficacy of UD heatmaps and the proposed algorithm in maximizing user connectivity as compared to K-means methods.

\end{abstract}


\section{Introduction}
Multi-Unmanned Aerial Vehicle (UAV) networks (MUNs) have emerged as a powerful paradigm, with coordinated operations that significantly enhance the versatility, scalability, and robustness of aerial missions\cite{shakeri2019design}. As a key component in beyond 5G communications, a MUN maneuverably enhances the service provisioning of cellular networks on-demand and fills the communication vacuum where terrestrial networks are not available\cite{geraci2022will}. However, attributed to UAVs' ad-hoc mobility, supporting ground communications with MUNs poses significant challenges, particularly for time-serial decision making over a long time horizon in dynamic, uncertain environments.

Recent advancements in deep reinforcement learning (DRL) have opened new avenues for addressing these challenges. With DRL, UAVs learn optimal collaborative strategies through trial-and-error interactions with the environments, enabling real-time adaptation to complex, time-varying conditions with minimal human intervention\cite{bai2023towards}. This is particularly crucial when UAVs must operate in unpredictable environments, such as disaster zones or hostile territories, where predefined strategies may be insufficient.

MUNs have been extensively studied in literature with centralized or distributed DRL. In centralized DRL, a master agent collects complete network information and makes decisions for all the UAVs\cite{ConsKha,luong2021deep,zhang2021learning}. For instance, Khairy \textit{et al.}\cite{ConsKha}, Luong \textit{et al.}\cite{luong2021deep} and Zhang \textit{et al.} \cite{zhang2021learning} studied joint altitude control and channel access management, joint UAV positioning and radio resource allocation, and user connectivity maximization in MUNs, using Proximal Policy Optimization (PPO), deep Q learning (DQL), and deep deterministic policy gradient (DDPG) algorithms, respectively. 
Distributed DRL, or multi-agent DRL, distributes the learning load across UAVs such that each UAV learns its own policy in a coordinated way to achieve overall objectives\cite{hu2021distributed,park2022cooperative}. For instance, Hu \textit{et al.} \cite{hu2021distributed} investigated trajectory design for a MUN to maximize ground user coverage. Multi-agent value decomposition based DRL was exploited. Park \textit{et al.} \cite{park2022cooperative} developed a collaborative multi-UAV positioning algorithm to achieve energy-efficient and reliable mobile access to cellular-vehicle communications. Multi-agent DQL was leveraged to tackle the optimization.

Nevertheless, most of the existing works on DRL-based MUN management assume a stationary user distribution (UD) \cite{ConsKha}\cite{park2022cooperative}\cite{li2024learning} or a dynamic one with predicted patterns\cite{zhang2021learning}\cite{tang2020deep}. In other words, the model of UD is a priori knowledge in the training, leading to UD-specific policies. When a MUN is deployed to unknown environments in occasions such as post-disaster management, ad-hoc activities, environment investigation, etc, such strategies may not perform well or require considerably extra training time to re-optimize. This makes optimal operations hardly available in a short time. Therefore, an adaptive MUN management strategy that can provide satisfying control for arbitrary UDs is desired. 

To this end, we study the design of DRL algorithms which can be generalized to arbitrary UDs after the training stage. A user connectivity maximization problem is considered where optimal UAV positioning is needed to maximize the number of served users by a MUN. To achieve generalization to arbitrary UDs, we propose to integrate Convolutional Neural Network (CNN) into deep Q network (DQN) to gain the ability of real-time UD analyzing. With the extracted high-level features from the CNN, optimal policies can then be determined. Specifically, our contributions are summarized as follows.
\begin{itemize}
\item The problem is first formulated into a time-coupled combinatorial nonlinear non-convex optimization. To make the optimization tractable and generalized to arbitrary UDs, a multi-agent CNN-enhanced DQL (MA-CDQL) algorithm is proposed leveraging on ResNet-based CNNs. The multi-agent framework is adopted to achieve scalability with the number of UAVs. 
\item To improve the exploration efficiency in learning and reduce the chance of converging to the local optimum, a heatmap algorithm is developed to transform the raw UD to a continuous density map. The generated heatmap will be part of the input to the policy network.
\item Simulations are conducted to demonstrate the efficacy of UD heatmap in avoiding local optimums and the MA-CDQL algorithm in achieved user connectivity compared to the K-means methods.
\end{itemize}

The reminder of the paper is organized as follows. Section \ref{sec.systemmodel} presents the system model and problem formulation. Section \ref{sec.hmap} explains the motivation and generation process of UD heatmap. Section \ref{sec.design} elaborates the design of the MA-CDQL algorithm. Section \ref{sec.simu} shows the numerical results, followed by the conclusion in Section \ref{sec.con}.

\section{System Model}\label{sec.systemmodel}
\begin{figure}[!ht]
	\centering
	\includegraphics[width=2.7in]{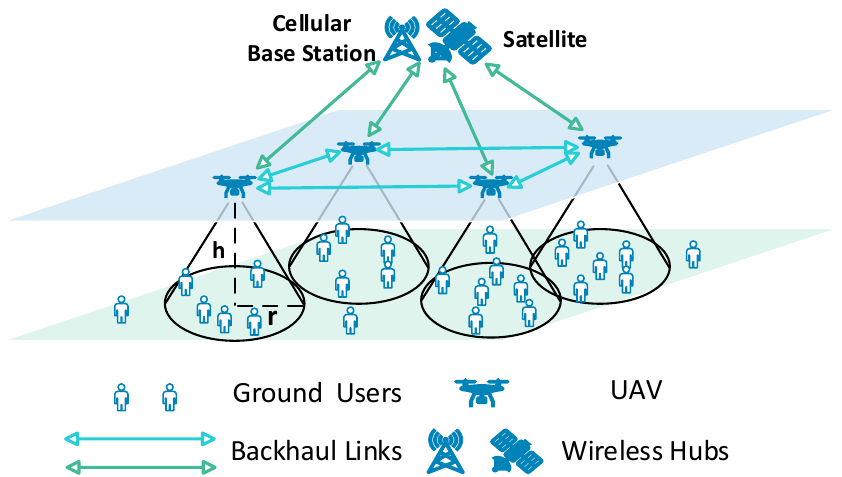}
    \caption{Network Model.} \label{fig.SystemModel}
\end{figure}

\subsection{Network Model}\label{subsec.NetMdl}
We consider a $L$-by-$L$ target area, across which a set of users, denoted as $\mathcal{U}$, are randomly distributed. Most of the users are clustered around hot spots while the remaining are scattered outside. To provide downlink communication services to the ground users, a group of UAVs, denoted as $\mathcal{I}$, are deployed, flying at a fixed altitude $H$ over the target area. Each UAV is equipped with a high-gain directional antenna that focuses most of its transmission power within a downward angle $\theta$, forming a circular communication footprint on the ground. The coverage radius can be expressed as $r = H \cdot \tan(\theta / 2)$. 

UAVs communicate with each other for coordination via wireless hubs (e.g., cellular base stations or a satellite). The UAV-hub links and UAV-user links occupy disjoint spectrum, thus causing no co-channel interference. Free of remote or centralized controllers, each UAV decides its own moves based on its local observations and mutually exchanged information. The network model is illustrated in Fig. \ref{fig.SystemModel}.
\subsection{Spectrum Access}\label{subsec.sa}
UAVs offer the same radio spectrum to ground users. Users access the shared spectrum via Orthogonal Frequency Division Multiple Access (OFDMA)\cite{mahmood2023joint}. With OFDMA, spectrum of each UAV is divided into orthogonal resource blocks (RBs), with a total number $N_{rb}$. Each UAV assigns different RBs to its users. Thus one user does not interfere with other users of the same UAV, but may suffer from interference from other UAVs that also cover it using overlapping RBs. We consider that each user has a minimum throughput requirement $r_{min}$. Therefore, if a user $u$ is connected to a UAV $i$, it will be randomly assigned with $N_{i,u}\leq N_{rb}$ RBs satisfying
\begin{equation}\label{eq.sa}
\sum^{N_{i,u}}_{k=1} W_{RB} \log_2(1+SINR_{i,u,m_k}) \geq r_{min},
\end{equation} 
where $W_{RB}$ denotes the bandwidth of one RB, $SINR_{i,u,m_k}$ denotes the signal-to-interference-and-noise ratio of user $u$ from UAV $i$ at RB $m_k$, and $m_k$ is the index of the $k$th random RB assigned to user $u$. $SINR_{i,u,m_k}$ is expressed as
\begin{equation}\label{eq.snir}
\begin{array}{l}
SINR_{i,u,m_k} = \frac{P_i G_{i,u,m_k}}{n_0 + \sum\limits_{{j \neq i},\;{j \in {\mathcal{I}}_{u,m_k}} } P_j G_{j,u,m_k}},\\ 
\text {where }G_{i,u,m_k}=10^{-PL_{i,u,m_k}/10}.
\end{array}
\end{equation} 
where $P_i$ and $P_j$ denote the transmit power spectrum density (PSD) of UAV $i$ and $j$, respectively, $G_{i,u,m_k}$ denotes the channel gain from UAV $i$ to user $u$ at RB $m_k$, $n_0$ denotes the PSD of noise, and $PL_{i,u,m_k}$ denotes path loss in dB from UAV $i$ to user $u$ at RB $m_k$. $\mathcal{I}_{u,m_k}$ denotes the set of UAVs that cover user $u$ and transmit over RB $m_k$. It may change with the UAV positions. A commonly adopted air-to-ground channel model \cite{al2014modeling} is exploited to estimate $PL_{i,u,m_k}$ as follows.
\begin{equation}\label{eq.pl}
PL_{i,u,m_k} = 20\log_{10}{d_{i,u}} + 20\log_{10}{f_{m_k}} - 27.55 + \eta \;\;\;  \text{(dB)},
\end{equation} 
where $f_{m_k}$ is the center frequency of RB $m_k$ in MHz, $d_{i,u}$ is 3D distance between UAV $i$ and user $u$, and $\eta$ represents excessive path loss related to Line-of-Sight (LoS) availability and propagation environments (e.g., urban or suburban).

\subsection{User Association}\label{subsec.userassl}
We consider a two-stage user association policy. In the first stage, each user determines the UAV that provides the best average SINR, and sends it a connection request. Once requests are collected, each UAV admits users according to a descending order of the reported SINR in the requests. Each UAV only admits a user when its available RBs satisfy Eq. \eqref{eq.sa}. Thus users making requests are not always guaranteed with admission. In the second stage, any unassociated user will request admission to its next best UAV (if any) until it is eventually admitted or all the UAVs are tried. Though not optimal, this policy can achieve a fast and near-optimal RB allocation. The proposed learning algorithm can also apply to any user association policy.

\subsection{Problem Formulation}
Our objective is to develop a trajectory control strategy to guide a set of UAVs to maximize the number of served users over a time horizon given \textit{arbitrary UDs}. The decision variables are each UAV's movements in each time step, constrained by the UAV spectrum availability and user association. The problem formulation is as follows.
\begin{equation}
\begin{array}{rl}
&\max_{a_{i,t},\;\forall i\in \mathcal{I}}\left[\sum_{t=1}^T\sum_{i \in \mathcal{I}}{\sum_{u \in \mathcal{U}}{X_{u, i}(t)}}\right] \text{\quad\quad\quad(P)}\\
\mbox{s.t.} & \quad X_{u, i}(t) \in \{0, 1\}, \forall{i \in \mathcal{I} \text{ and } \forall u \in \mathcal{U}} \text{\quad\quad\quad\quad(C1)} \\
&\quad \sum_{i \in \mathcal{I}}{X_{u, i}(t)} \leq 1 , \; \forall{\: u \in \mathcal{U}} \text{\quad\quad\quad\quad\quad\quad\quad\;\;(C2)}\\
&\quad\sum_{u \in {\mathcal{U}}}{N_{i,u}} \leq N_{rb}, \forall{i \in \mathcal{I}} \text{\quad\quad\quad\quad\quad\quad\quad\quad\;(C3)}\\
&\quad 0 \leq x_{i,t}, \;y_{i,t} \leq L, \forall{i \in \mathcal{I}} \text{\quad\quad\quad\quad\quad\quad\quad\quad\;\;(C4)}
\end{array}\nonumber
\end{equation}
In problem P, timing of the network is slotted into $T$ steps, indexed by $t$. At step $t$, $a_{i,t}$ denotes the moving direction and distance of UAV $i$, and $(x_{i,t},y_{i,t})$ denote its horizontal position. $X_{u,i}(t)$ is a binary indicator taking 1 when user $u$ is associated with UAV $i$ at step $t$, and 0 otherwise. $\{X_{u,i}(t)\}$ are jointly determined by UAV positions, UAV RB allocation and user association. Constraint $C_2$ requires each user associated to at most one UAV at each step. Constraint $C_3$ ensures that the available RBs of each UAV are not over assigned. Constraint $C_4$ limits the UAVs within the target region.

When $\{a_{i,t}\}$ are discrete and finite, problem P is a time-coupled combinatorial nonlinear non-convex optimization problem. What is more challenging is that for different UDs, the optimization problem needs to be retackled to obtain the new optimum. This is prohibitively complicated for onsite operation where computing resources are limited and UAVs need to be settled in a short time. Instead of solving the optimization directly, our proposed method utilizes CNN to extract the high-level features of the detected UD and outputs via offline trainings an adaptive strategy which can infer optimal UAV positions for arbitrary UDs with low computing complexity.


\section{Generation of User Distribution Heatmap}\label{sec.hmap}
\begin{wrapfigure}{r}{0.2\textwidth}
    \centering
    \includegraphics[width=1.0\linewidth]{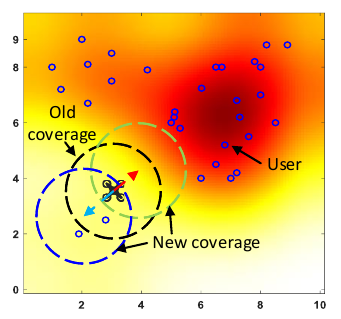}
    \caption{\small{Motivation illustration of UD heatmap}}
    \label{fig.illu}
\end{wrapfigure}

The motivation of generating UD heatmap is illustrated in Fig. \ref{fig.illu}. When a UAV is at a position with sparsely distributed users (e.g., the black coverage), its movement in the middle way towards hot spots (e.g., to the green coverage) may bring no change or even a decrease in the number of served users. Such situations act negatively in helping the UAV find the optimal position. Instead of using the discrete distribution, a heatmap can be generated reflecting the DENSITY of UD in a continuous manner. When the UAV moves towards the hot spots, its covered density increases even if its covered number of users reduces. This will effectively increase the exploration efficiency and reduce the chance of being trapped in the local optimum.

The process of heatmap generation begins from a scattered-point UD (e.g., image $S_0$ in Fig. \ref{fig:sg}). It consists of two stages: heatmap transformation, and heatmap smoothing. Specifically, the entire area is treated as an $N$x$N$ meshgrid. Each grid intersection is a pixel in the heatmap. The algorithm first calculates the number of users within a radius of $coh\_ r$ to each pixel to get a coarse heatmap (e.g., $H_0$ in Fig. \ref{fig:sg}). The map is then smoothed into a more continuous one (e.g., $H_1$) by averaging neighboring pixels within a square range of $broad\_r$ for $Itr$ iterations. The details are shown in Algorithm \ref{alg:HMgen}.

\begin{algorithm}
\caption{Generation Algorithm of UD Heatmap}\label{alg:HMgen}
\begin{algorithmic}[1]
\small{\REQUIRE User position list $U$, Coherent radius $coh\_r$, Smooth range $broad\_r$, Iterations $Itr$;
\ENSURE Heatmap $H$ of size $N$x$N$;
\STATE \textbf{/*Heatmap generation*/}
\FOR{$i, j = 1$ to $N$}
    \STATE Initialize $pix\_val = 0$;
    \STATE Get grid intersection point $G(i,j)$;
    \FOR{each user $k$}
        \IF{user $k$ is within $coh\_r$ of $G(i,j)$}
            \STATE $pix\_val = pix\_val + 1$;
        \ENDIF
    \ENDFOR
    \STATE $H_{i,j} = pix\_val$;
\ENDFOR
\STATE \textbf{/*Heatmap smoothing*/}
\FOR{$it = 1$ to $Itr$}
    \FOR{$i, j = 1$ to $N$}
        \STATE Set smoothing boundaries:
        \STATE $up = \min(j + broad\_r, N)$;
        \STATE $dn = \max(j - broad\_r, 1)$;
        \STATE $lt = \max(i - broad\_r, 1)$;
        \STATE $rt = \min(i + broad\_r, N)$;
        \STATE Compute the average of the submatrix $H(lt:rt, dn:up)$:
        \STATE $H'_{i,j} = \frac{1}{(rt-lt+1)(up-dn+1)} \sum{H(lt:rt, dn:up)}$;
    \ENDFOR
    \STATE Update $H \gets H'$;
\ENDFOR
\RETURN $H$}
\end{algorithmic}
\end{algorithm}

\section{Design of Multi-Agent CNN-Enhanced Deep Q-Learning (MA-CDQL) Algorithms}\label{sec.design}
In this section, we present the design details of the MA-CDQL algorithm based on the obtained UD heatmap. 
The multi-agent framework requires the UAVs to exchange information for coordination. In our design, each UAV will share its position and number of served users at each step.

\subsection{State Definition} 
The target area is discretized into an $M$x$M$ grid, with UAVs taking positions only at the grid intersections. The individual state space of each UAV needs to include \textit{i)} information of UD for generalization to arbitrary UDs, and \textit{ii)} positions of other UAVs for coordination. We define the state space as a 3D matrix to be analyzable by the CNN. The 3D matrix is composed of three 2D matrices of the same size. The first 2D matrix is the $M$x$M$ heatmap (e.g., $H_2$ in Fig. \ref{fig:sg}) downsampled from the $N$x$N$ heatmap. Downsampling aligns the heatmap to the other 2 2D matrices in size and reduces the state space and the training complexity. The second 2D matrix is the local UAV position matrix, denoted as $\mathcal{PL}^k$ for UAV $k$, with its element $\mathcal{PL}^k_{i,j}$ defined as
\begin{equation}\label{eq.local}
\mathcal{PL}^k_{i,j}=\left\{
\begin{array}{rl}
1,& \text{if UAV $k$ is at grid intersection (i,j)}\\
0,& \text{otherwise.}
\end{array} \right. 
\end{equation} 
The third 2D matrix is the global UAV position matrix, denoted as $\mathcal{PG}^k$ for UAV $k$, with its element $\mathcal{PG}^k_{i,j}$ defined as
\begin{equation}\label{eq.global}
\mathcal{PG}^k_{i,j}=
n, \text{ if $n$ UAVs are at grid intersection (i,j)}.
\end{equation} 
Note that $\mathcal{PL}^k$ and $\mathcal{PG}^k$ are not combined into one matrix because it is advantageous to differentiate the position of the local UAV from those of the other UAVs. In this way, each UAV knows where itself is relative to others so that it can better determine where to go. Denote the individual state of UAV $i$ as $s_i$, and $s_i$ can be expressed as a 3D matrix obtained by stacking the three 2D matrices together, i.e.,
\begin{equation}\label{eq.state2}
s_{i} = (H_2|\;\mathcal{PL}^{i}|\;\mathcal{PG}^{i}),\;\forall i\in\mathcal{I}.
\end{equation}

\subsection{Action Definition} 
When UAV positions are limited only to the grid intersections, each UAV has only 5 possible moves at each intersection, i.e., move forward, backward, to the left, to the right, and hover still. When it moves, it will move by a constant distance from one intersection to an adjacent one. Denote the individual action space of UAV $i$ as \(\mathcal{A}_i\), which can be expressed as
\begin{equation}\label{eqn8}
\mathcal{A}_i = \{0, 1, 2, 3, 4\},
\end{equation} 
where 0,1,2,3,4 corresponds to the above 5 actions.
\subsection{Reward Function}
The reward function for one individual UAV is composed of four parts: \textit{a)} the number of its served users, \textit{b)} the total number of served users by all the UAVs, \textit{c)} the total heatmap density covered by all the UAVs, and \textit{d)} the distance penalty for coverage overlapping with other UAVs. 

Part a) of the reward is defined in Eq. \eqref{localnum}. It guides each UAV to serve as many users as possible, directing the UAV toward areas with denser UDs.
\begin{equation}\label{localnum}
R^{local\_ num}_{i}(t) = \sum_{u \in \mathcal{U}} {X_{u,i}(t)} , \: \forall \; {i \; \in \; \mathcal{I}}.
\end{equation}  
However, relying solely on a) may lead to uncoordinated UAV positioning, where some UAVs occupy dense regions early and casually, preventing others from finding good positions. To tackle this, UAVs may exchange information of their respective number of served users to collaboratively maximize the total number of served users, i.e., part b). Part b) is defined as
\begin{equation}\label{eqn10}
\begin{split}
R^{glb\_ num}_{i}(t) &= \sum\limits_{u \in \mathcal{U},\; j \in \mathcal{I}} X_{u,j}(t)/{|\mathcal{I}|}, \: \forall \: i \: \in \: \mathcal{I}.
\end{split}
\end{equation} 

Part c) is to give the UAV a hint on if they are potentially heading towards hot spots at one step. This is to compensate the negative effect illustrated in Fig. \ref{fig.illu} that sometimes a good move may result in decreased part a) and b) but increased part c). The part c) reward is calculated as follows. Define an $M$x$M$ coverage matrix $\mathcal{HC}(t)$ at step $t$, with each element corresponding to a grid intersection and calculated as
\begin{equation}\label{eq.cvg}
\mathcal{HC}_{i,j}(t)=\left\{
\begin{array}{rl}
1,& \text{if any UAV covers intersection ($i$,$j$)};\\
0,& \text{otherwise}
\end{array} \right. 
\end{equation} 
Then part c), i.e., the global density reward, is obtained as:
\begin{equation}\label{eq.partc}
\begin{split}
R^{glb\_ dens}_{i}(t) &= \sum_{i=1}^{M} \sum_{j=1}^{M} (H_2)_{i,j}(t)\cdot\mathcal{HC}_{i,j}(t)  
\end{split}
\end{equation}

Last but not least, part d) is to punish coverage overlapping between UAVs such that UAVs are well scattered to reduce contention. It is calculated according to the following equation.
\begin{equation}\label{eq.penalty}
\begin{array}{l}
R_i^{cvg\_ pnt}(t)=\sum\limits_{j \in \mathcal{I}, j\ne i} \max\left(0, \;(1-\frac{d_{i,j}(t)}{2r}) \cdot p_{max}\right), \\
\text{where } p_{max} = d_p \cdot \frac{|\mathcal{U}|}{|\mathcal{I}|}.
\end{array}
\end{equation}
In Eq. \eqref{eq.penalty}, \(r, \;d_{i,j}(t),\;p_{max}\) denotes the ground coverage radius of a UAV, the distance between UAV $i$ and $j$, and the maximum distance-based penalty, respectively. The value $d_p$ is an adjustable weighting factor that tunes the impact of $R_i^{cvg\_ pnt}(t)$ on the total individual reward $R_i(t)$. When one UAV is at least $2r$ distance away from all the other UAVs, it will not get penalized, i.e., $R_i^{cvg\_ pnt}(t)=0$. 

The total individual reward $R_i(t)$ is calculated as a weighted summation of all the parts defined above, as given in Eq. \eqref{eq.totalreward}, where $\alpha_1$, $\alpha_2$ and $\alpha_3$ are weighting constants.
\begin{equation}\label{eq.totalreward}
\begin{array}{ll}
R_{i}(t)=&\alpha_{1}*R_i^{local\_num}(t)+\alpha_{2}*R_i^{glb\_num}(t)+\\&\alpha_{3}*R_i^{glb\_dens}(t) + R_i^{cvg\_pnt}.
\end{array}
\end{equation}

\subsection{Neural Network Design} 
\begin{figure}[h]
    \centering
    \begin{subfigure}[b]{0.45\textwidth}
        \centering
        \includegraphics[width=0.85\textwidth]{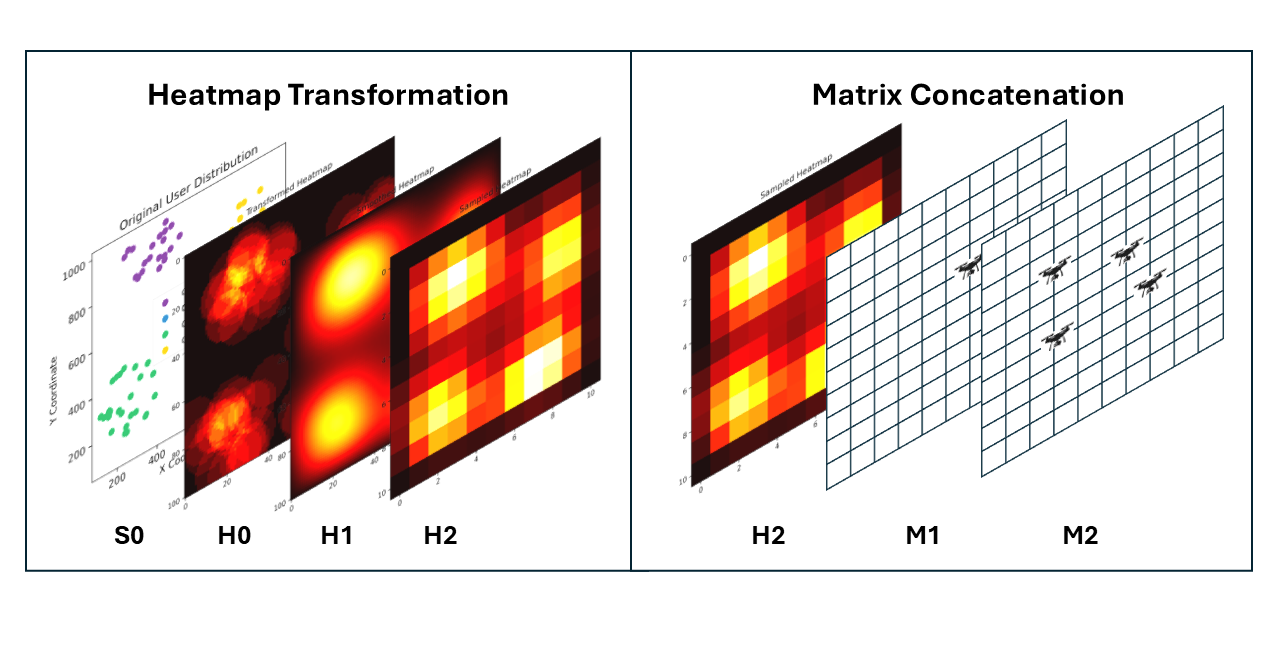}
        \caption{UD heatmap and state generation}
        \label{fig:sg}
    \end{subfigure}
    \begin{subfigure}[b]{0.45\textwidth}
        \centering
        \includegraphics[width=0.9\textwidth]{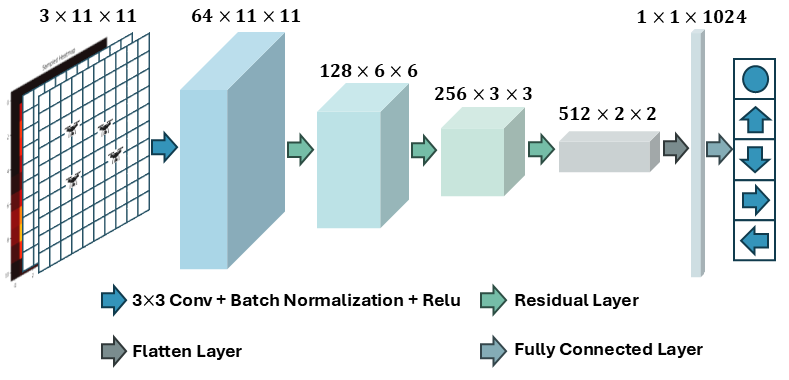}
        \caption{The structure of the CNN-enhanced deep Q network.}
        \label{fig:cnn}
    \end{subfigure}
    \caption{State generation and the structure of CDQN.}
    \label{fig:overview}
\end{figure}

The architecture of the CDQN is based on ResNet\cite{he2016deep}. The input state matrix is first saved, then processed through convolution, batch normalization and ReLu activation into a 64-channel feature map. The feature map is then passed through 3 residual layers. Each residual layer is composed of 6 residual blocks which further consists of 2 convolutional layers and a residual connection. The results are then flattened into a one-dimension vector, with the action output generated through a fully connected layer. 
The input matrix and architecture of the CDQN are shown in Fig. \ref{fig:cnn}.
\subsection{Implementation} 
The MA-CDQL algorithm consists of three phases: UD heatmap generation, model training, and algorithm execution. In the first phase, a pool of UDs are randomly generated where most users are clustered in random hot spots and the remaining are randomly distributed across the target area. Each UD is transformed into a UD heatmap according to Algorithm I and downsampled. The procedure is shown in Fig. \ref{fig:sg}. During the training phase, in each episode, the environment is initialized with a UD from the pool and the corresponding downsampled UD heatmap. Throughout the training, UAVs interact with the environment, exchange information, and update their respective Q networks (as shown in Fig. \ref{fig:cnn}) until final convergence. During the execution phase, when deployed in a scene with an unknown UD, UAVs will first cooperatively sweep the area to get the scattered-point UD, from which downsampled UD heatmap is obtained. Based on the downsampled heatmap and the real-time positions of all the UAVs, each UAV is guided by the distributed policy to its optimal position that collectively maximizes the total number of served users.
\section{Experiment}\label{sec.simu}
\subsection{Simulation Setup}
The target area is discretized into a $10 \times 10$ grid. Each grid cell has a side length of 100 meters. The training was conducted in Windows 10 with a 16-core vCPU and an RTX 4090 GPU. The entire experiment was implemented in Python using the PyTorch library. Training was performed for up to 30,000 episodes, each having 30 time steps. Three different settings were considered in the experiment. All the settings have 100 users. Setting 1 has 3 UAVs and 3 hot spots, with two hot spots having significantly more users than the third; 10 users are outside the hot spots. Setting 2 has 3 UAVs and 4 hot spots, with each hot spot being clearly separated; 10 users are outside the hot spots. Setting 3 has 5 UAVs and 4 hot spots, with all the users in the hot spots. The main parameters are summarized in table \ref{tab:simparam}.

\begin{table}[!ht]
\footnotesize
\centering
\caption{Simulation Parameters}

\renewcommand{\arraystretch}{1}

\begin{tabular}{!{\vrule width0.8pt}l|l!{\vrule width0.8pt}}\Xhline{0.8bp}
\multicolumn{1}{!{\vrule width0.8pt}c|}{\gape{\bfseries Parameters}} & \multicolumn{1}{c!{\vrule width0.8pt}}{\gape{\bfseries Values}} \\ 
         \hline
         \rowcolor[gray]{0.9}
         UD in hot spots: Setting 1 & (15, 35, 40)\\
         UD in hot spots: Setting 2 & (20, 20, 25, 25)\\
         \rowcolor[gray]{0.9}
         UD in hot spots: Setting 3 & (25, 20, 30, 25) \\
         Number of RBs per UAV, RB bandwidth $W_{RB}$ & {20, 180$kHz$} \\
         \rowcolor[gray]{0.9}
         {Altitude $H$} and aperture angle $\theta$ & {$350m$ and $60^{\circ}$} \\
         Spectrum center frequency $f_c$ & 2GHz\\
         \rowcolor[gray]{0.9}
         Transmit and noise psd: ($P_i, n_0$)& (-49.5, -174)dBm\\ 
         Min. user throughput $r_{min}$ & 250kbps\\
         \rowcolor[gray]{0.9}
         LoS path loss parameter $\eta$ & 1dB\\
         CDQN learning rate & 3.5e-4\\
         \rowcolor[gray]{0.9}
         {Epsilon $\epsilon$, Discount factor $\gamma$, Mini-batch size} & {0.1, 0.95, 128} \\
         Target network updating rate & 1/10 \\
         \rowcolor[gray]{0.9}
         Reward weighting factor ($\alpha_1$,$\alpha_2$,$\alpha_3$,$d_p$) & (1, 0.5, 1, 3)\\
\hline
\end{tabular}
\label{tab:simparam}
\end{table}

\subsection{Simulation Results}
\begin{figure}[!ht]
	\centering
	\includegraphics[width=2.6in]{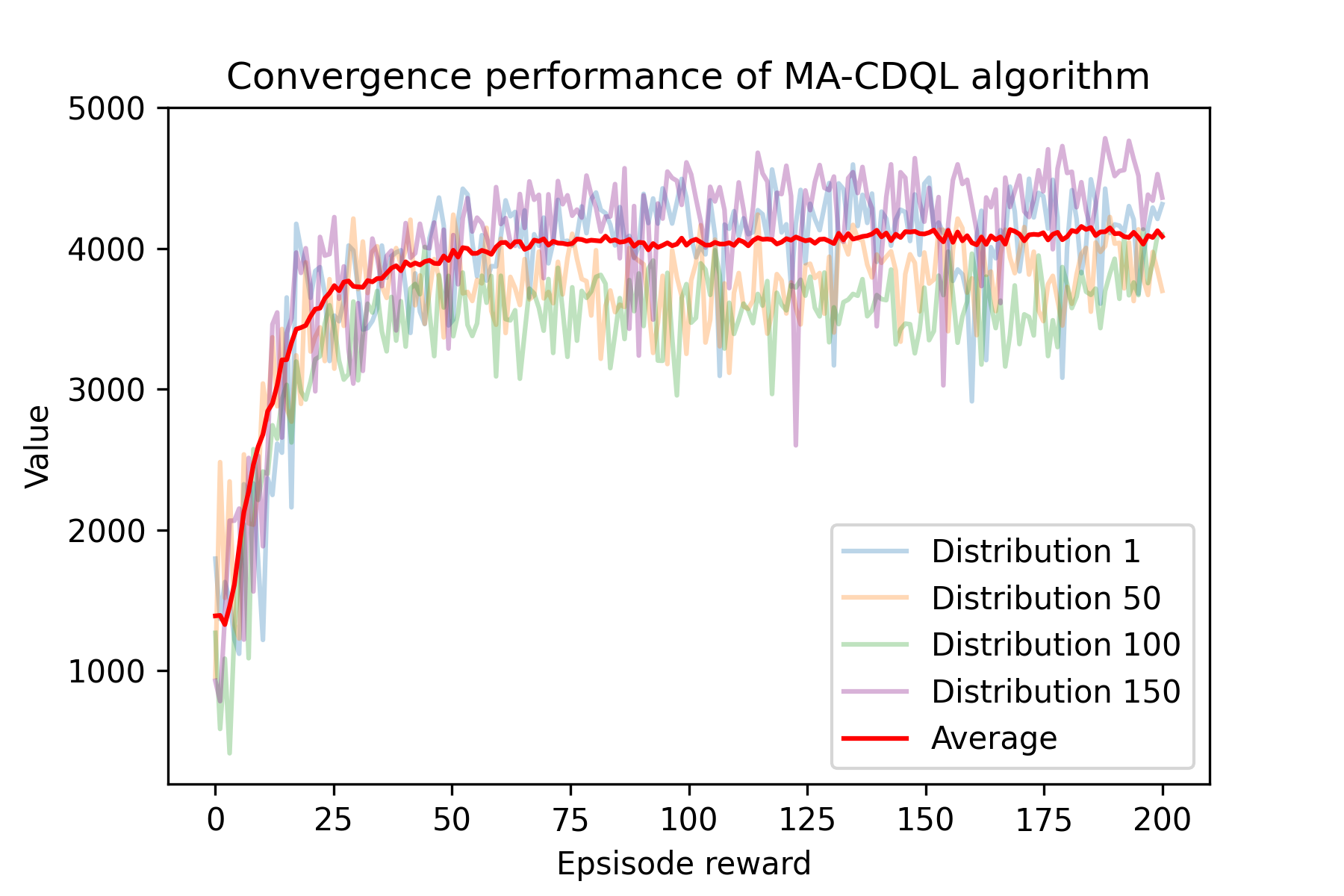}
    \caption{Convergence performance of MA-CDQL algorithm} \label{fig.convergence}
\end{figure}

We first test the convergence performance of the MA-CDQL algorithm, and present the results in Fig. \ref{fig.convergence}. 150 randomly generated UDs (75 for Setting 1 and 2, respectively) are used, each of which is initialized into the environment every 150 episodes. Thus in Fig. \ref{fig.convergence}, we group every 150 episodes into an epoch. We randomly selected several distributions to display the individual convergence, along with the average convergence over all distributions. It can be seen that the overall training converges well.

We then compare user connectivity performance between three methods: K-means clustering\cite{park2024k}, the MA-CDQL w/o heatmap, and MA-CDQL w/ heatmap. Algorithms in K-means class can quickly identify user clusters with low computing complexity, but it is challenging to integrate more complicated constraints (e.g., user throughput requirement, spectrum availability) for better performance. For MA-CDQL without heatmap, it calculates the number of users in each grid and shapes the numbers into an $M$x$M$ matrix as part of the state space. Its reward function considers both individual and overall number of served users. The results are shown in table \ref{tab:results}.

In the table, \textbf{trS} and \textbf{teC} refers to the training and testing set of a specific setting, respectively. The table shows that MA-CDQL yields considerably better user connectivity in Setting 1 and 2 than K-means, and the heatmap version outperforms the non-heatmap version in all settings. Yet in Setting 3, K-means achieves better performance than the proposed method. 

\begin{table}[h]
\centering
\caption{Performance Comparison}
\label{tab:results}
\begin{tabular}{|c|c|c|c|c|c|c|}
\hline
\textbf{} & \textbf{trS1} & \textbf{trS2} & \textbf{trS3} & \textbf{teS1} & \textbf{teS2} & \textbf{teS3} \\
\hline
\textbf{kmeans} & 48.53 & 48.80 & 86.49 & 49.06 & 48.76 & \textbf{86.55} \\
\hline
\textbf{non-HM} & 53.77 & 54.23 & 83.09 & 52.61 & 54.08 & 80.05 \\
\hline
\textbf{w/ HM} & \textbf{56.96} & \textbf{58.65} & \textbf{87.34} & \textbf{55.64} & \textbf{57.71} & 83.36 \\
\hline
\end{tabular}
\end{table}

\begin{figure}[h]
    \centering
    \begin{subfigure}[b]{0.2\textwidth}
        \centering
        \includegraphics[width=0.95\textwidth,height=1.2in]{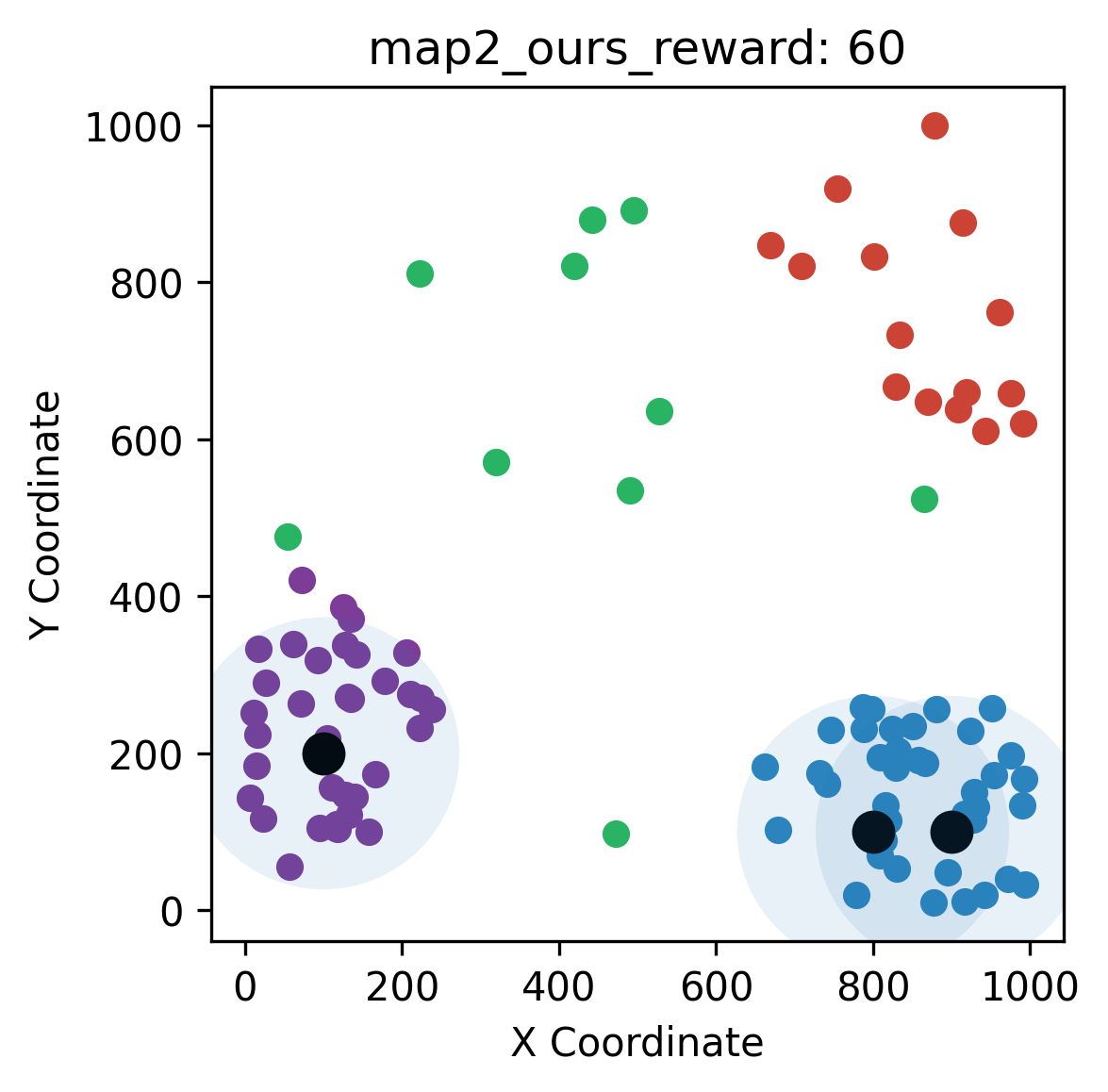}
        \caption{Setting1\_MA-CDQL}
        \label{fig:image3}
    \end{subfigure}
    \begin{subfigure}[b]{0.2\textwidth}
        \centering
        \includegraphics[width=0.95\textwidth,height=1.2in]{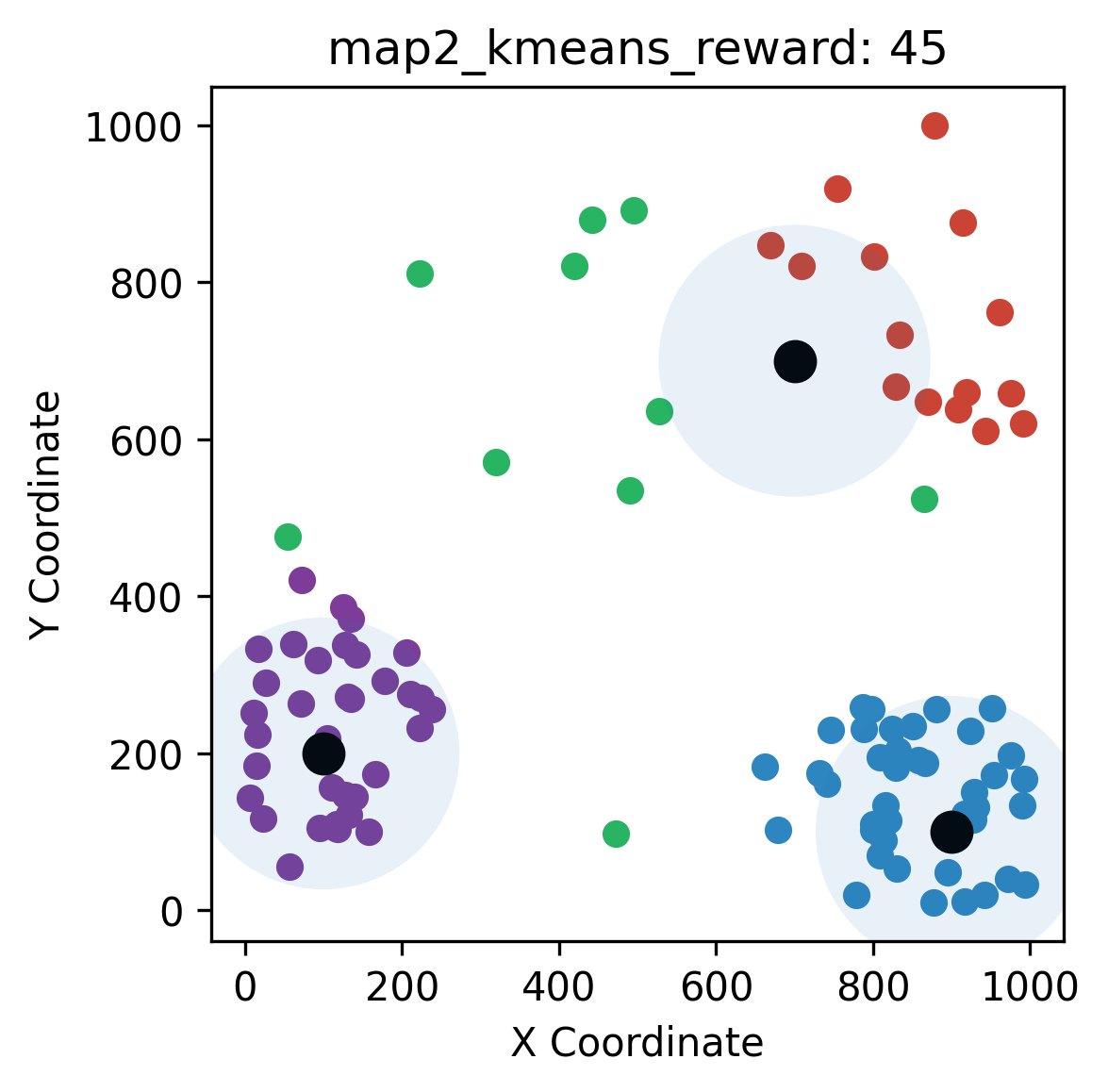}
        \caption{Setting1\_K-means}
        \label{fig:image4}
    \end{subfigure}
    \begin{subfigure}[b]{0.2\textwidth}
        \centering
        \includegraphics[width=0.95\textwidth,height=1.2in]{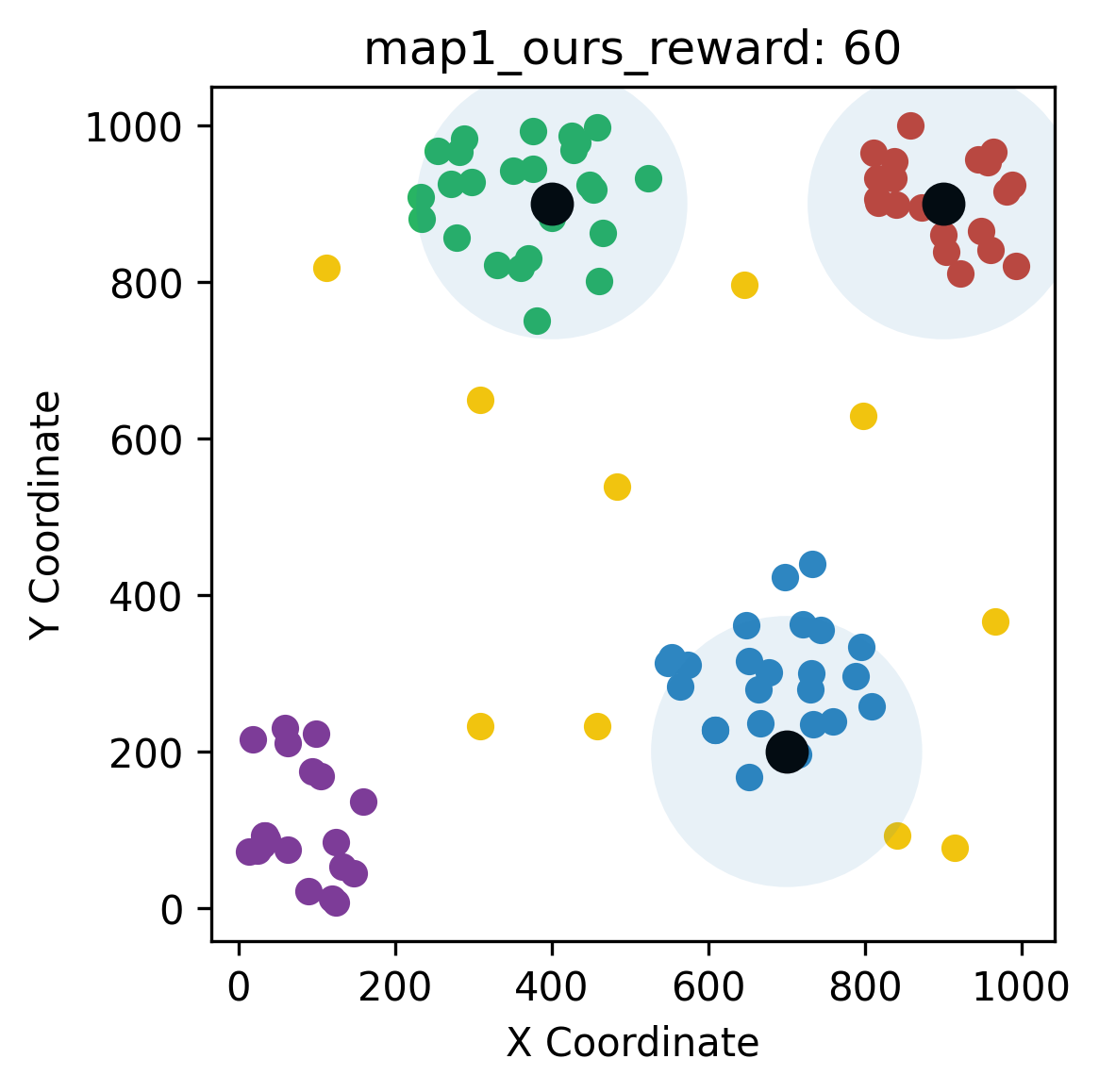}
        \caption{Setting2\_MA-CDQL}
        \label{fig:image1}
    \end{subfigure}
    \begin{subfigure}[b]{0.2\textwidth}
        \centering
        \includegraphics[width=0.95\textwidth,height=1.2in]{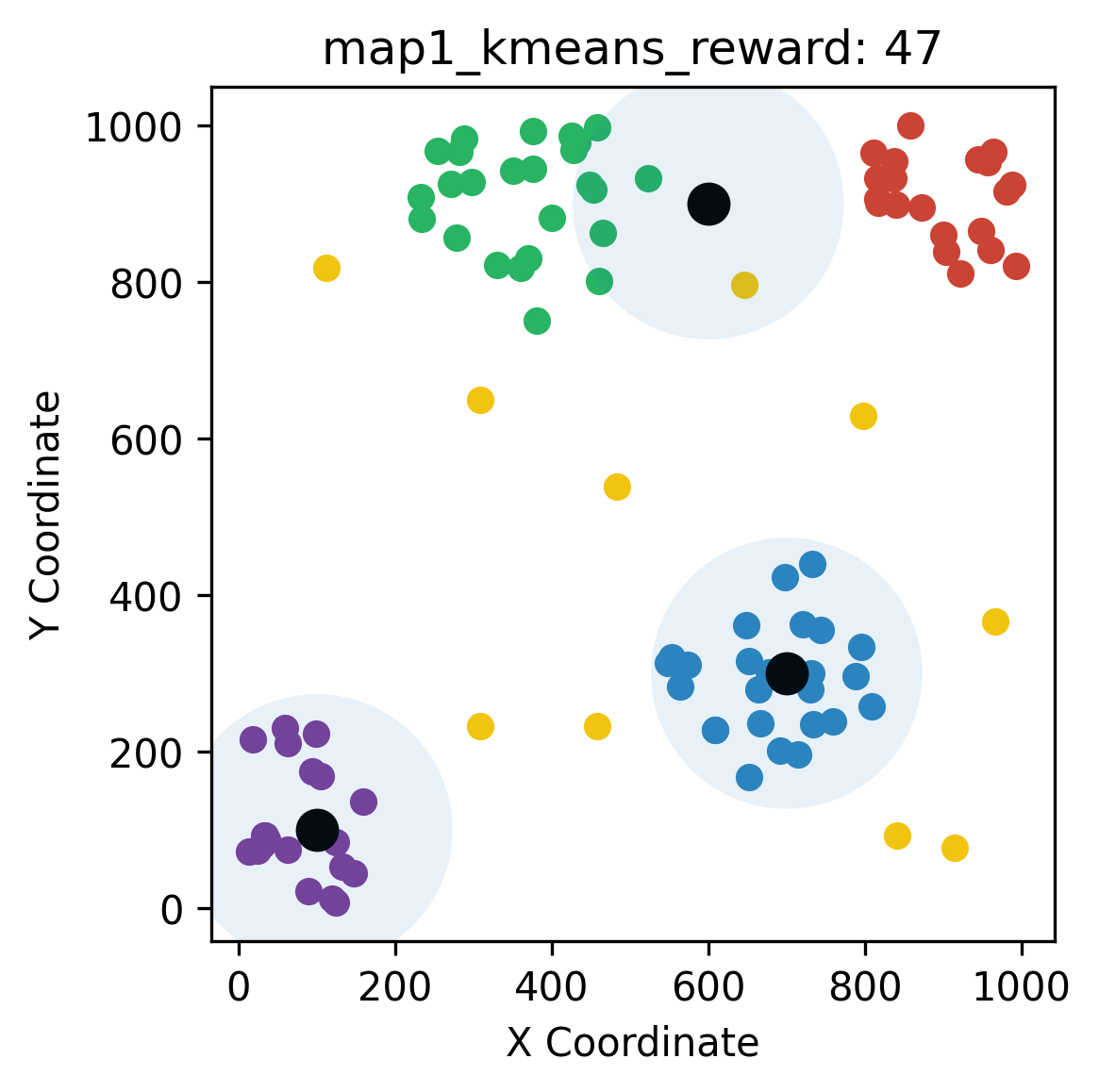}
        \caption{Setting2\_K-means}
        \label{fig:image2}
    \end{subfigure}

    \caption{Comparison between MA-CDQL and K-means.}
    \label{fig:four_images}
\end{figure}

To provide insights of the performance gain in Setting 1 and 2, 2 example UDs and UAV deployments are given in Fig. \ref{fig:four_images}. In Setting 1, the blue and purple hot spots have much more users than the red one. K-means cannot integrate UAV spectrum availability which determines the maximum number of users a UAV can serve. As a result, it deploys only 1 UAV to each cloud and ends up with 25$\%$ fewer users than our proposed method. In Setting 2, K-means requires to determine the number of clusters before execution. Since the number of hot spots is unknown for a random UD, the number of clusters will be set to 3 by default. As a result, it combines the green and red hot spots into one and deploys the UAV in the middle, where there are fewer users. The proposed method does not have the cluster concept and is user-amount-oriented, thus yielding 27$\%$ more users than K-means.

Therefore, it can be concluded that \textit{i}) compared to K-means, our proposed method does not always outperform, especially in cases when the number of UAVs exceeds the number of hot spots; \textit{ii}) however, the proposed method yields better generalization across various types of clustering distributions with considerably better or comparable user connectivity performance; \textit{iii}) integration of heatmap can effectively improve the user connectivity performance in all settings. 

\section{Conclusions}\label{sec.con}
This paper has studied distributed user connectivity maximization problem with generalization to arbitrary user distributions (UDs). To make the problem tractable, the MA-CDQL algorithm has been proposed. The algorithm has integrated a ResNet-based CNN into the deep Q network to extract the high-level features of the input UD and infer optimal guidance to UAVs to maximize user connectivity. A UD heatmap algorithm has been developed to smooth the UD into continuous density map in order to improve the learning efficiency and avoid local optimums. Simulations have demonstrated that \textit{i)} compared to K-means methods, the proposed method yields better generalization across various UD clustering settings with considerably better or comparable user connectivity performance; and \textit{ii)} integration of UD heatmap can effectively further improve the performance in all settings.

\bibliographystyle{IEEEtran}
\bibliography{main}

\end{document}